\documentclass[12pt,a4paper]{article}
\usepackage{a4wide}
\usepackage{latexsym}
\usepackage{epsf}
\usepackage{amssymb}
\usepackage[english]{babel}
\usepackage{amsfonts}
\usepackage{amsmath}
\usepackage{fancyhdr}
\usepackage{floatflt,graphicx}
\usepackage{psfrag}
\usepackage{cite}
\begin{document}
\def\be{\begin{equation}}
\def\ee{\end{equation}}
\def\bea{\begin{eqnarray}}
\def\eea{\end{eqnarray}}
\def\pd{\partial}
\def\a{\alpha}
\def\b{\beta}
\def\g{\gamma}
\def\d{\delta}
\def\m{\mu}
\def\n{\nu}
\newcommand{\fsl}{{\hspace{-7pt}\slash}}
\newcommand{\gsl}{{\hspace{-5pt}\slash}}
\newcommand{\dslash}{\pd\fsl}
\newcommand{\pslash}{p\gsl }
\newcommand{\Aslash}{A\gsl }
\newcommand{\kslash}{k\gsl }
\newcommand{\Lslash}{\Lambda\gsl }
\newcommand{\kuslash}{k_1\gsl }
\newcommand{\kdslash}{k_2\gsl }
\newcommand{\Dslash}{D\fsl}
\def \h{\mathcal{H}}
\def \hh{\mathcal{G}}
\def\t{\tau}
\def\p{\pi}
\def\th{\theta}
\def\l{\lambda}
\def\O{\Omega}
\def\r{\rho}
\def\s{\sigma}
\def\e{\epsilon}
  \def\scri{\mathcal{J}}
\def\cM{\mathcal{M}}
\def\tcM{\tilde{\mathcal{M}}}
\def\RR{\mathbb{R}}
\def\tr{\operatorname{tr}}
\def\str{\operatorname{str}}
\begin{flushright}
IFT-UAM/CSIC-06-31\\
hep-th/0606267\\
\end{flushright}
\vspace{1cm}
\begin{center}
{\bf\Large  Renormalized masses of heavy Kaluza-Klein states.}
\vspace{.5cm}
{\bf Enrique \'Alvarez and Ant\'on F. Faedo }
\vspace{.3cm}
\vskip 0.4cm  
 
{\it  Instituto de F\'{\i}sica Te\'orica UAM/CSIC, C-XVI,
and  Departamento de F\'{\i}sica Te\'orica, C-XI,\\
  Universidad Aut\'onoma de Madrid 
  E-28049-Madrid, Spain }
\vskip 0.2cm
\vskip 1cm
{\bf Abstract}
\par
\end{center}
\begin{quote}
Several ways of computing the radiative corrections to the heavy boson masses in
Kaluza-Klein theory are discussed. It is argued that only an intrinsically higher 
dimensional approach embodies all the desired physical properties. This contradicts earlier
results in the literature.
  
\end{quote}
\newpage
\setcounter{page}{1}
\setcounter{footnote}{0}
\tableofcontents
\newpage
\section{Introduction}
The vacuum polarization  of a gauge theory is a valuable source of information. 
It conveys information on  the running of the corresponding gauge  coupling and 
in principle it can be used to compute would-be radiative corrections to the mass 
of the gauge bossons. These corrections vanish in the usual four dimensional theory 
owing to unbroken gauge invariance; that is, the Ward-Takahashi identity. However, this last 
statement cannot be directly applied to a theory defined in dimensions greater than 
four because of its peculiarities, in particular, the presence of an infinite tower 
of Kaluza-Klein states from the four dimensional point of view.
This has motivated a vast number of studies on these issues, including the 
possibility of a power law behaviour of the couplings \cite{Dienes, Oliver} 
and finiteness of the radiative Higgs mass in Gauge-Higgs unification models 
\cite{Antoniadis, Hatanaka}. In this models the Higgs is identified with the extra 
components of a gauge field in higher dimensions. 
We want to focus our attention on the calculation of the radiative mass of 
the extra dimensional gauge boson with trivial holonomy, i.e., we will not
 consider noncontractible Wilson loops.
\par
The physical intuition behind these ideas is that higher dimensional gauge invariance
somewhat protects the Higgs from getting radiative contributions to its mass. And
for
this to be true, it is plain that at very short distances, physics must be really higher-
dimensional.
\par
There are essentially three different ways to compute these corrections. We shall comment on 
them in turn, and argue eventually that if we want to formally implement the aforementioned
 ideas, a full higher dimensional computation is mandatory.
 
The correction has been often computed diagramatically once the mode expansion and the 
integral over the compact manifold had been performed, which means that in some sense this 
computation is purely four-dimensional because the Feynman rules applied correspond to 
a theory with an infinite number of Kaluza-Klein (KK) modes and their corresponding 
interactions (see 
for example \cite{Cheng, vonGersdorff}). The result of this kind of calculations 
is a one loop finite mass for the Higgs field proportinal to the compactification scale.
The main purpose of this paper is to repeat the calculation from a 
different point of view based on the discussion done in \cite{Alvarez}, wherein 
 a systematic method for computing directly in higher dimensions is introduced, and, as
we have already said, it is 
claimed that this is crucial because it is the only one that actually implements
the physical intuition that at very short distances all dimensions should become visible,
and at any rate, it does not give the same results as a purely four-dimensional calculation.
Let us indeed analyze the divergences of a quantum field theory using 't Hooft's ideas 
\cite{ 'tHooft} applied to higher dimensions (which we shall always take to be either $n=5$,
 or $n=6$ for the sake of the argument)
We shall expand all fields \footnote{Although we are doing this for bosonic fields
only, it can be easily generalized to fermions with only minor modifications, along
the lines of 't Hooft's original paper.}
around an arbitrary background, $\bar{\phi}$,
\be\label{fondo}
\Phi_i=\bar{\phi}_i+\phi_i
\ee
(where the subindex stands for spacetime as well as internal degrees of freedom).
The expansion of the action up to quadratic order in the fields is
\be\label{general}
i S= i\int d^n x \left[S(\bar{\phi}_k)+(\frac{\delta 
S}{\delta\Phi_l}(\bar{\phi})+ J_l)\phi_l+
J_l \bar{\phi}_l+
\phi_i(-\frac{1}{2}\Box\delta_{ij}-N_{ij}^{\m}(\bar{\phi})\pd_{\m}
-\frac{1}{2}M_{ij}(\bar{\phi}))\phi_j\right]
\ee
The term linear in the fluctuations is absent whenever the background is a solution
of the equations of motion, which we will assume from here on.
The partition function is given by the gaussian integral
\be
Z(\bar{\phi},J)\equiv\int{\cal D}\phi e^{i S}=e^{i [S(\bar{\phi})+
\int d^n x J_l\bar{\phi}_l]}
\det\,^{-1/2}[{\cal M}_{ij}]
\ee
where ${\cal M}_{ij}(\bar{\phi})\equiv -\frac{1}{2}\Box
\delta_{ij}-
N_{ij}^{\m}(\bar{\phi})\pd_{\m}
-\frac{1}{2}M_{ij}(\bar{\phi})$.
It follows that
\be
 W(\bar{\phi},J)\equiv -i \log Z(\bar{\phi},J)=S(\bar{\phi})+\int d^n x 
J_l\bar{\phi}_l
+\frac{i}{2}\log\det[{\cal M}_{ij}(\bar{\phi})]
\ee
The piece involving the determinant reads
\be
\frac{i}{2}\log\, \det [\frac{i}{2}\Box]+\frac{i}{2}\log\, 
\det [\delta_{ij}+\Box^{-1}
(2N^{\m}_{ij}\pd_{\m}+M_{ij})]=\frac{i}{2}\sum_{n=0}^{\infty}
\frac{(-1)^{n+1}}{n} 
\tr [\Box^{-1}(2N^{\m}_{ij}\pd_{\m}+M_{ij})]^n
\ee
The last equality indicates the way of computing the determinant using Feynman diagrams.
In four dimensions, the counterterm has got dimension four. 
The most general counterterm of mass dimension four is (taking into account that 
 $[M]=2$ y $[N]=1$)
\bea
\Delta L&=&\frac{1}{8\pi^2\epsilon}\tr[a_0 M^2+a_1 (\pd_{\m}N_{\n})^2+
a_2 (\pd_{\m}N^{\m})^2+\nonumber\\
&&a_3 M N_{\a}N^{\a}+a_4 N_{\m}N_{\n}\pd^{\m}N^{\n} +a_5 (N_{\a}N^{\a})^2+
a_6 (N_{\m}N_{\n})^2 ]
\eea
In order to compute the coefficients, it is enough to compute a few selected diagrams.
Before doing that, and also before studying the appropiate extension to five and six 
dimensions, it is convenient to recall a hidden symmetry again uncovered by 't Hooft.
\par
Let us rewrite the lagrangian in a compact notation as
\be
L=\frac{1}{2}(\pd_{\m}\phi+N_{\m}\phi)^2-\frac{1}{2}\phi X \phi
\ee
with
\be
X\equiv M-N_{\a}N^{\a}
\ee
There is now a manifest $O(N)$ invariance
\bea
\delta\phi&=&\Lambda(x)\phi(x)\nonumber\\
\delta N_{\m}&=&-\pd_{\m}\Lambda+[\Lambda,N_{\m}]\nonumber\\
\delta X &=&[\Lambda,X]
\eea
The one-loop counterterm must respect this symmetry. The most general Lorentz invariant,
dimension four operator with this property reads
\be
\Delta L_{n=4}=\frac{1}{8\pi^2\epsilon}\tr [a X^2 + b F_{\m\n}F^{\m\n}]
\ee
where $F_{\m\n}\equiv\pd_{\m}N_{\n}-\pd_{\n}N_{\m}+[N_{\m},N_{\n}]$.
\par
Explicit computation yields
\bea 
a&=&\frac{1}{4}\nonumber\\
b&=&\frac{1}{24}
\eea
The whole of the preceding reasoning goes through to six dimensions, and
actually to any {\em even} dimension. The most general counterterm (before using the background 
equations of motion ) is given by
\bea
\Delta L_{n=6}&=&\frac{1}{8\pi^2\epsilon}\tr \left[a\, X^3  +b\, D_\a F_{\b\gamma}
D^\a F^{\b\gamma}+c\, X F_{\m\n}F^{\m\n}  + d\, D_\a X D^\a X\right.\nonumber\\
&&\left.+\,e F_{\m\n}F^{\n\rho}F_{\rho}^{\m}+f\,D^2 D^2 X + 
g\, D_{\a}F^{\a\b}D^{\gamma}F_{\gamma\b}\right]
\eea
Again, computation of a few diagrams fully determine the numerical coefficients. More
efficient techniques based upon the heat kernel can also be applied (cf. \cite{Alvarez}).
\par
This same reasoning gives no candidate counterterms in five dimensions (nor in 
any {\em odd} dimension). What happens is the following. When using the proper time
(explained, for example, in Collins' book \cite{Collins}) representation of a propagator
\be
\frac{1}{(p^2-m^2)^a}=\int_0^{\infty}d\tau\, \tau^{a-1} e^{-\tau (p^2-m^2)}
\ee
after using Feynman parameters in an arbitrary dimension, one ends up with
integrals of the type
\be
I\equiv \int_0^1 dx \int_0^{\infty}d\tau\, \tau \left(\frac{\pi}{-\t}\right)^{n/2} 
e^{-\tau k^2 x(1-x)}  
\ee
The integral over proper time is then done, yielding a Gamma function
\be
\Gamma(2-n/2)
\ee
which has poles only for even values of the dimension. This integral over proper time involves
an analytic continuation. The divergence can be isolated by imposing a cutoff in the lower limit
of the integral (this is {\em not} a cutoff in momentum space, and is compatible with 
whatever symmetries the theory enjoys), and expanding the integrand in powers of $\t$ 
\be
I\equiv \int_0^1 dx \int_{\Lambda^{-2}}^{\infty}d\tau\, \tau \left(\frac{\pi}{-\t}\right)^{n/2} 
(1-\tau k^2 x(1-x)+O(\t^2))  
\ee
Using this procedure, we get in four dimensions a logarithmic divergence; just another language
to express the divergence previously studied; symbollically,
\be
\frac{1}{\e}\sim \log\frac{\Lambda}{\m},
\ee
$\m$ being an infrared cutoff.
The difference is that this gives in five dimensions a nontrivial result, namely a
linear divergence. We can then write:
\be
\Delta L_{n=5}=\Lambda_{n=5}\, \tr [a\, X^2 + b\, F_{\m\n}F^{\m\n}]
\ee
In six dimensions this gives both a quadratic and a logarithmic divergence. The general 
structure of the six-dimensional counterterm would then be
\bea
\Delta L_{n=6}&=&\Lambda_{n=6}^2\, \tr [a\, X^2 + b\, F_{\m\n}F^{\m\n}]+\nonumber\\
&&\log\frac{\Lambda_{n=6}}{\m_{n=6}}\tr \left[c\, X^3  +d\, D_\a F_{\b\gamma}
D^\a F^{\b\gamma}+e\, X F_{\m\n}F^{\m\n}  + f\, D_\a X D^\a X\right.\nonumber\\
&&\left.+ g\, F_{\m\n}F^{\n\rho}F_{\rho}^{\m}+h\, D^2 D^2 X +  
i D_{\a}F^{\a\b}D^{\gamma}F_{\gamma\b}\right]
\eea
We can now try to make precise the physical intuition that tells us that a five dimensional
theory in $\mathbb{R}^4\times S^1$ whould become four-dimensional in the limit in which
the Kaluza-Klein scale $M$ (the inverse of the radius of the circle) gets much bigger
than any other scale. In this limit we can approach any five dimensional integral by
\be
\int d^5 x f(x^\m,x_4)\sim \frac{1}{M}\int d^4 x \bar{f}(x^\m)
\ee
(where $x^\m$ are the usual four-dimensional coordinates). It is then possible (and necessary 
for mathematical 
consistency of the physical intuition) to choose the five and four dimensional cutoffs 
in such a way that
\be
\frac{\Lambda_{n=5}}{M}=\log\,\frac{\Lambda_{n=4}}{\m_{n=4}}
\ee
We can then contemplate a chain of reductions from six dimensions to five 
dimensions (at a scale $M_6$) and from five to four (at a scale $M_5$)\footnote{A
 somewhat similar 
analysis is done in the heat kernel language starting from supergravity in eleven dimensions
by Fradkin and Tseytlin in \cite{Fradkin}.}
\bea
\frac{\Lambda_{n=6}^2}{M_6 }&=&\Lambda_{n=5}\nonumber\\
\frac{\Lambda_{n=5}}{M_5}&=&\log\,\frac{\Lambda_{n=4}}{\m_{n=4}}
\eea
The six-dimensional logarithmic divergence appears then in four dimensions as a $\log\,\log$
divergence. 
\par
The preceding ideas lead to a general procedure to renormalize theories in a way consistent
with dimensional reduction \cite{Alvarez}. Let us examine its consequences in a couple 
of examples, namely, a version of $QED_{n=5}$ and $QED_{n=6}$.
\section{Six-dimensional vacuum polarization} 
As we show in the Appendix, there are inherent ambiguities in a four dimensional calculation 
and we will try to avoid them by computing directly in the higher dimensional space. 
Suppose we have $QED_6$, quantum electrodynamics on a six-dimensional manifold. The theory is 
of course non renormalizable because the coupling constant has mass dimension $[e]=-1$. 
Nevertheless it is possible to identify and study all divergences appearing at one loop 
order (or $O(e^2)$)\cite{Alvarez}.
The one loop divergences are given,  for a flat manifold and in terms of 
the gauge background field $\bar{A}_M$, and the fermionic backgrounds $\bar{\eta}, \eta$
\begin{eqnarray}\label{sdcounter}
\Delta L_{n=6}&=&\Lambda_{n=6}^2\int\frac{d^6 x}{(4\pi)^3}\left(\frac{4}{3}e^2\bar{F}_{MN}^2+
4 e^2 \bar{\eta}\bar{\Dslash}\eta
+12me^2\bar{\eta}\eta\right)+\nonumber\\
&&\log\frac{\Lambda_{n=6}}{\m_{n=6}}\int\frac{d^6 x}{(4\pi)^3}\left(-\frac{4}{3}
e^2m^2\bar{F}_{MN}\bar{F}^{MN}-2e^2m^2\bar{\eta}
\gamma^M\bar{D}_M\eta-6e^2m^3\bar{\eta}\eta\right.-\nonumber\\
& &-\left.\frac{11}{45}e^2\bar{D}_R\bar{F}_{MN}\bar{D}^R\bar{F}^{MN}+
\frac{23}{9}e^2\bar{D}_M\bar{F}^{MN}\bar{D}^R\bar{F}_{RN}+\frac{19}{15}e^2m\bar{\eta}
\bar{D}_M\bar{D}^M\eta\right)+\nonumber\\&&
+O(e^3)
\end{eqnarray}
In addition to the counterterms corresponding to operators that were already present in the 
original Lagrangian  higher order operators have been generated radiatively. The appearance
 or this terms was discussed in \cite{Ghilencea, Oliver}. If we want to absorb their divergences 
we must include them in the bare Lagrangian 
\be\label{bare}
\mathcal{L}=\mathcal{L}_0+\m D_MF^{MN}D^RF_{RN}+\l D_RF_{MN}D^RF^{MN}+\dots
\ee
Where $[\m]=[\l]=-2$. We have written explicitly only the extra terms that are cuadratic 
in the gauge field and therefore the ones that modify the extra-dimensional vacuum polarization.
 Once we perform the mode expansion the same operators will yield the mass of the tower coming 
from the gauge field. If we define
\be
A_M^0=Z_3^{1/2}A_M
\ee
We get 
\be
Z_3=1+\frac{e^2}{12\p^3}\Lambda^2_{(d=6)}-\frac{e^2m^2}{12\p^3}\log{\frac{\Lambda^2_{(d=6)}}
{\m^2_{(d=6)}}}.
\ee
 It is easy to see then that the pole in $F_{MN}^2$ 
is absorbed in the wave function renormalization of the gauge field so from an 
extra dimensional point of view there is no renormalization of the mass of the gauge 
boson\footnote{This can be seen by introducing a mass term $m_B^2A_MA^M$ in the bare 
Lagrangian. Repeating the computation of \cite{Alvarez} it is found that the only 
effect at one loop is the apearance of extra fermionic terms like $m_B^2e^2\left(\bar{\psi}
\Dslash\psi+m_f\bar{\psi}\psi\right)$ so $m_B$ remains unrenormalized}. 
This is expected in some sense due to gauge invariance. In four dimensions it is well 
known that even if we include a mass term for the photon in the bare Lagrangian its mass 
does not renormalize. 
Nevertheless, gauge invariance is not enough to ensure a massless photon as we know from the
 Schwinger model in two dimensions. The lesson to learn from this is that the number of 
dimensions is crucial. Since in Gauge-Higgs unification the Higgs bosson is identified 
with the extra-dimensional components of the gauge field once the mode expansion has been 
performed its mass does not renormalize either.
Concerning higher order terms, its divergences can be absorbed in arbitrary dimensionful
 couplings like $\m$ and $\l$ in (\ref{bare}) if we define
\be
\begin{array}{c}
\m_0=Z_\m\m\\
\l_0=Z_\l\l
\end{array}
\ee
The conclusion is the very same as for $F_{MN}^2$: once we have renormalized the theory 
in six dimensions the mass coming from the mode expansions does not renormalize because 
the divergences are absorbed in $Z_3$ and $Z_\m$,$Z_\l$. Of course, to all orders of 
perturbation theory we would need an infinite number of arbitrary couplings to fit with 
experiments and this is preciselly the benchmark for a non renormalizable field theory.
\par
It is interesting to study the effects of this extra operators at tree level. First of all they 
induce corrections to the mass of the gauge bossons once the compactification has been 
performed.
 For example in six dimensions compactification of (\ref{bare}) yields terms like
\be\label{cormas}
(\m+2\l)|N|^4A_\m^{-n}A_n^\m
\ee
 And similar ones (i.e. of order $(2\l+\m)M^4$) for the scalar field. Observe that at one 
loop we find a renormalization of the 
dimensionful couplings $\m$ and $\l$ that induces a running for the masses 
through (\ref{cormas}) 
which is suppressed by $M^{-2}$ (with respect to the usual mass). 
Concerning the propagator suppose now that we include higher order terms in the form
\be
F_{MN}^2+\frac{c_1}{\Lambda^2}F_{MN}\pd^2F^{MN}+\frac{c_2}{\Lambda^2}F_{MN}\pd^M\pd_RF^{RN}
\ee
Where $\Lambda$ is a parameter (naturally of the order of the compactification mass) in 
order to make $c_1$ and $c_2$ dimensionless. Then the propagator of the gauge field is
\be
A_M D_{MN}^{-1}A_N=A_M\left(1-\frac{2c_1+c_2}{\Lambda^2}p^2\right)\left(p^2\delta_{MN}-
p_Mp_N\right)A_N
\ee
It has the usual pole in $p=0$, but also depending on the sign of the couplings $c_1$ and 
$c_2$ it can have another one
\be
p^2\sim\frac{M^2}{2c_1+c_2}
\ee
It may be possible to use arguments \cite{Adams} concerning superluminal fluctuations 
around non-trivial backgrounds to fix the sign of the couplings and avoid this second pole. 
In any case possible poles coming from this higher order terms can be absorbed in 
dimensionful coupling constants introduced in the bare Lagrangian in the form (\ref{bare}).
 Therefore, in some sense, the mass of the gauge field is protected from renormalization.
In this respect, it is interesting to point out Smilga's conjecture \cite{Smilga} that
there might exist consistent higher derivative theories, in particular in six dimensions
(although the zero mode instability is always present).
\section{Five-dimensional vacuum polarization}
Let us turn our attention to five-dimensional QED,
 on  $\mathbb{R}^4\times S^1$ 
\begin{equation}
S=\int d^5x\left(\frac{1}{4}F_{MN}^2+\bar{\psi}^i\left(\Dslash_{ij}+m_f\d_{ij}\right)\psi^j
\right)
\end{equation}
We have  doubled the fermion content of the theory and defined new matrices ($i,j=1,2$)
\be
\g^M_{ij}\equiv\g^M\otimes\s^2_{ij}
\ee
that satisfy a modified Clifford algebra
\be\label{ca}
\{\g^M,\g^N\}_{ij}=2\d^{MN}\otimes\d_{ij}
\ee
for computational reasons.
Standard computations lead to the counterterm
\be\label{a4ext}
\Delta L_{n=5}=\Lambda_{n=5}\int\frac{d^5x}{(4\pi)^{5/2}}\left(\frac{4}{3}e^2\bar{F}_{MN}^2+
3e^2\bar{\eta}^i
\bar{\Dslash}_{ij}\eta^j+10e^2m_f\bar{\eta}^i\eta^i\right)
\end{equation}
There are some differences with respect to the six-dimensional theory of the previous section.
To be specific, there is no logarithmic divergence, neither are higher dimensional
operators  generated as counterterms (they start to appear at two-loop order)
\footnote{Dimensional regularization analysis yields a finite answer from a five-dimensional 
point of view. This is a well known result common to odd dimensional spacetimes at 
one loop order,
as we already mentioned in the introduction.}.
This result disagrees with the one   obtained from a four dimensional analysis
even if one takes into account the whole KK tower (which will be reviewed in the Appendix,
cf.   the formula
(\ref{a4tow}))
 unless we find some 
consistent definition of the
infinite sums in such a way that each one vanishes.
In particular, as in six dimensions, gauge invariance forbids a mass term for the gauge 
boson. Therefore
the masses of the modes once we make the expansion are protected because we can absorb 
the divergence 
in $Z_3$ as we argued in the last section. Explicitly
\be
Z_3=1+\frac{16e^2}{3(4\p)^{5/2}}\Lambda_{n=5}
\ee
This is not the case from the four-dimensional point of view, 
as we see from (\ref{a4tow}). A detailed description of the renormalization from the point 
of view 
of four dimensions is given in the Appendix, where we point out  the ambiguities 
and inconsistencies
of that choice.
\section{Conclusions}
We have analyzed radiative corrections in extra-dimensional theories not involving gravity.
It has been shown that a four dimensional calculation is at least ambiguous when one considers
 the theory at one loop. There are two different ways of computing diagrams according 
to the place
 where the mode sum over all the KK tower is performed. When the sum is done 
after the momentum integral 
(which corresponds
 to the calculation of the Appendix  with the whole tower) usual four-dimensional divergences 
are found
 along with extra divergences coming from infinite sums. Also we find many problems
 with the divergence of the mass of the zero mode scalar because it is massless at tree level.
If one adopts, as we advocate, the higher dimensional point of view with the purpose of 
renormalizing the 
theory then the possible counterterms are dictated by gauge and Lorentz invariance in the 
extra dimensional manifold. This fixes the form of the possible mass terms for the 
four-dimensional gauge bosson as well as the Higgs in Gauge-Higgs unification. 
Therefore, it is easy to convince oneself that every divergence may be absorbed in the 
wave function renormalization of the gauge background $\bar{A}_M$ and the renormalization 
of the couplings
of higher dimension operators 
such as $\m$ and $\l$ in (\ref{bare}).
\par
This approach is, in our opinion, the only one that embodies the physical intuition, which we
believe correct, that at very short distances all dimensions should appear at the same foot,
and physics should be higher dimensional.
\section*{Acknowledgments}
This work has been partially supported by the
European Commission (HPRN-CT-200-00148) and by FPA2003-04597 (DGI del MCyT, Spain), as well as
 Proyecto HEPHACOS (CAM); P-ESP-00346.      
A.F. Faedo has been supported by a MEC grant, AP-2004-0921. We are indebted to Bel\'en Gavela
 for useful discussions and to R. Contino, M. Quir\'os and A. Santamar\'ia for illuminating correspondence.
\newpage
\appendix
\section{Four-dimensional vacuum polarization corresponding to the KK tower.}
In this Appendix we will review the diagrammatical four-dimensional calculation in order to 
illustrate its inherent difficulties. We will follow closely the computation done in 
\cite{Cheng} but performing the sum over the extra dimensional momentum at the end. 
Consider the vacuum polarization function of $QED_5$. If one of the dimensions corresponds 
to a circle $S^1$ with radius $R$ then the 
momentum in that dimensions is quantized in units of $R^{-1}\equiv M$ and the integral has 
to be replaced by a sum. Taking into account the Feynman rules the vacuum polarization
 has the form ($p^2=p_\m p^\m$) 
\bea
&&i\Pi_{\m\n}(p^2,p_5^2)=-e^2\sum_{k_5}\int \frac{d^4k}{(2\p)^4}\tr
\left(\g_\m\frac{1}{\kslash+i\g^5k_5}\g_\n\frac{1}{\left(\kslash-\pslash\right)+
i\g^5\left(k_5-p_5\right)}\right)=\nonumber\\&=&
-4e^2\sum_{k_5}\int\frac{d^4k}{(2\p)^4}\frac{k_\m\left(k_\n-p_\n\right)+
k_\n\left(k_\m-p_\m\right)-g_{\m\n}k\left(k-p\right)+g_{\m\n}k_5\left(k_5-p_5\right)}
{\left(k^2-k_5^2\right)\left(\left(k-p\right)^2-\left(k_5-p_5\right)^2\right)}
\eea
Introducing a Feynman parameter and doing the usual shift in the four-momentum 
$k_\m'=k_\m-\a p_\m$ as well as a shift in the compact dimension $k_5'=k_5-\a p_5$ we get
\be
i\Pi_{\m\n}=-4e^2\sum_{k_5}\int^1_0d\a\int\frac{d^4k}{(2\p)^4}\frac{N_{\m\n}}
{\left(k^2-k_5'^2+\a\left(1-\a\right)\left(p^2-p_5^2\right)\right)^2}
\ee
Where the numerator is
\be
N_{\m\n}=2k_\m k_\n+g_{\m\n}\left(-k^2+\a\left(1-\a\right)\left(p^2-p_5^2\right)+
\left(2\a-1\right)p_5k'_5+k'^2_5\right)-2\a\left(1-\a\right)p_\m p_\n
\ee
And we have neglected terms linear in $k_\m$ which vanish because of the 
angular integral. Let us then split the vacuum polarization into two pieces.
\be
\Pi_{\m\n}\equiv g_{\m\n}\Pi_1-p_\m p_\n\Pi_2
\ee
After Wick rotation to Euclidean space
\begin{displaymath}
\Pi_1=-4e^2\sum_{k_5}\int^1_0d\a\int\frac{d^4k}{(2\p)^4}\frac{\frac{k^2}{2}+
\a\left(\a-1\right)\left(p^2+p_5^2\right)+\left(2\a-1\right)p_5k'_5+k'^2_5}
{\left(k^2+k_5'^2+\a \left(1-\a \right)\left(p^2+p_5^2\right)\right)^2}
\end{displaymath}
\be
\Pi_2=8e^2\sum_{k_5}\int_0^1d\a\left(1-\a\right)\a\int\frac{d^4k}{(2\p)^4}\frac{1}
{\left(k^2+k_5'^2+\a\left(1-\a\right)\left(p^2+p_5^2\right)\right)^2}
\ee
Using a proper time parametrization the first piece can be put into the form
\bea
\Pi_1&=&-4e^2\sum_{k_5}\int^1_0d\a\int_0^\infty d\t \t\int\frac{d^4k}{(2\p)^4}
\left(\frac{k^2}{2}+\a\left(\a-1\right)\left(p^2+p^2_5\right)+\left(2\a-1\right)
p_5k'_5+k'^2_5\right)\times\nonumber\\&&
\times e^{-\t\left(k^2+k_5'^2+\a\left(\a-1\right)\left(p^2+p^2_5\right)\right)}
\eea
The integral in momentum space is obviously cuadratically divergent, but it can be
computed in dimensional regularization:
\bea
\Pi_1&=&-\frac{4e^2\p^{n/2}}{(2\p)^n}\sum_{k_5}\int^1_0d\a\int_0^\infty d\t 
\t\left(\frac{n}{4\t^{\frac{n}{2}+1}}+\frac{\a\left(\a-1\right)\left(p^2+p^2_5\right)+
\left(2\a-1\right)p_5k'_5+k'^2_5}{\t^{\frac{n}{2}}}\right)\times\nonumber\\&&
\times e^{-\t\left(k_5'^2+\a\left(\a-1\right)\left(p^2+p^2_5\right)\right)}
\eea
It is now easy to perform the integral in proper time and particularize to $n=4+\e$ 
dimensions to get
\bea\label{diag}
\Pi_1&=&-\frac{4e^2\p^{n/2}}{(2\p)^n}\Gamma\left(2-\frac{n}{2}\right)\sum_{k_5}
\int^1_0d\a\left(\frac{n}{4\left(1-\frac{n}{2}\right)}\left(k_5'^2+\a
\left(\a-1\right)\left(p^2+p^2_5\right)\right)^{\frac{n}{2}-1}+\right.\nonumber\\
&+&\left.\left(\a\left(\a-1\right)\left(p^2+p^2_5\right)+\left(2\a-1\right)p_5k'_5+k'^2_5
\right)\left(k_5'^2+\a\left(1-\a\right)\left(p^2+p^2_5\right)\right)^{\frac{n}{2}-2}\right)=
\nonumber\\
&=&\frac{e^2}{12\p^2}\Gamma\left(-\frac{\e}{2}\right)\left(p^2+p_5^2+\frac{1}{2}p_5^2
\right)\sum_{k_5}1
\eea
Similar manipulations with $\Pi_2$ yield
\be
\Pi_{\m\n}(p^2,p_5^2)=\frac{e^2}{12\p^2}\Gamma\left(-\frac{\e}{2}\right)\left(\left(
p^2+p_5^2+\frac{1}{2}p_5^2\right)g_{\m\n}-p_\m p_\n\right)\sum_{k_5}1
\ee
Note that the vacuum polarization of the four-dimensional photon $A_\m^{(0)}$ (which 
means $p_5=0$) verifies the Ward-Takahashi identity 
\be
p^\m\Pi_{\m\n}(p^2,p_5=0)=0
\ee
From a four dimensional point of view this result is not surprising at all. For fixed 
$k_5$ it corresponds to the contribution to the pole of a single fermionic loop. 
If we now consider an infinite number of fermions coupled with the same strength
 to the gauge bosons we have an additional
 divergence coming from the sum over the whole tower. The  heat kernel computation done
 with the tower of modes in four dimensions seems to support this conclusion. The corresponding 
counterterm is
\bea\label{a4tow}
\Delta L_{n=4}&=&\int\frac{d^4x}{(4\pi)^2}\sum_l\left(\frac{4}{3}e^2\sum_n\bar{F}_{\m\n}^n
\bar{F}_{-n}^{\m\n}-4e^2\sum_n\bar{A}_5^{-n}\Box\bar{A}_5^n-16ie\frac{l}{R}
\left(\frac{l^2}{R^2}+m_f^2\right)\bar{A}_5^0+\right.\nonumber\\&&
\left.+4e^2\sum_n\left(2m_f^2+\frac{2n^2+(n+l)^2}{R^2}\right)\bar{A}_5^{n-l}
\bar{A}_5^{l-n}+4ie^3\sum_{n,m}\frac{2m+l+n}{R}\bar{A}_5^{m-l}\bar{A}_5^{l-n}
\bar{A}_5^{n-m}-\right.\nonumber\\&&\left.
-4e^4\sum_{n,m,s}\bar{A}_5^{m-l}\bar{A}_5^{l-s}\bar{A}_5^{s-n}\bar{A}_5^{n-m}+
8ie^2\sum_n\frac{n}{R}\pd_\m\bar{A}_5^n\bar{A}_{-n}^\m+4e^2\sum_n\frac{n^2}{R^2}
\bar{A}^\m_n\bar{A}_\m^{-n}+\right.\nonumber\\&&\left.
+6e^2\sum_{n\ne0}\bar{\eta}_{l-n}^i\dslash_{ij}\eta_{l-n}^j-6e^3\sum_{n,m\ne0}
\bar{\eta}_{l-m}^i\Aslash^{l-n}_{ij}\eta_{n-m}^j-12e^3\sum_{n,m\ne0}
\bar{\eta}_{l-m}^i\g^5_{ij}\bar{A}_5^{l-n}\eta_{n-m}^j+\right.\nonumber\\&&\left.
+12i\sum_{n\ne0}\frac{l}{R}\bar{\eta}_{l-n}^i\g^5_{ij}\eta_{l-n}^j+
20m_f\sum_{n\ne0}\bar{\eta}_{l-n}^i\eta_{l-n}^i+3e^2\bar{\eta}_l^i\dslash_{ij}
\eta_l^j-3e^3\sum_{n}\bar{\eta}_{n}^i\Aslash^{n-l}_{ij}\eta_{l}^j-\right.\nonumber\\
&&\left.
-6e^3\sum_{n}\bar{\eta}_{n}^i\g^5_{ij}\bar{A}_5^{n-l}\eta_{l}^j+6i\frac{l}{R}
\bar{\eta}_{l}^i\g^5_{ij}\eta_{l}^j+10m_f\bar{\eta}_{l}^i\eta_{l}^i+\right.\nonumber\\
&&\left.\left(6\frac{l^4}{R^4}-8m_f^2\frac{l^2}{R^2}-4m_f^4\right)\right)
\log{\frac{\Lambda_{n=4}}{\m}}
\eea 
From (\ref{a4tow}), if we define the renormalized field and mass 
($m_n^2=\frac{n^2}{R^2}$)
\be
\begin{array}{c}
A_{\m(0)}^n=Z_3^{1/2}A_\m^n\\
m_{n(0)}^2=Z_mm_n^2
\end{array}
\ee
Then we get
\be
\begin{array}{c}
Z_3=1+\frac{e^2}{3\p^2\e}\sum_l1\\
Z_m=1+\frac{e^2}{6\p^2\e}\sum_l1
\end{array}
\ee
There is an obvious divergence due to the infinite sum, which could be regularized, 
for example, by using a zeta function. Our results are, however, independent of the
particular definition chosen. 
With the renormalization group functions
\be
\begin{array}{c}
\b_e\equiv\frac{\pd e}{\pd\log\m}=\frac{e^3}{12\p^2}\sum_l1\\
\b_{m_n}\equiv\frac{\pd m_n}{\pd\log\m}=-\frac{e^2m_n}{12\p^2}\sum_l1
\end{array}
\ee
Notice that the beta function of the fine structure constant embodies an infinite 
number of identical fermion  contributions. The behavior of the couplings is
\be
\begin{array}{c}
e^2=\frac{e_0^2}{1-\frac{e_0^2}{6\p^2}\sum_l1\log{\frac{\m}{\m_0}}}\\
m_n=m_n^0\left(1-\frac{e_0^2}{6\p^2}\sum_l1\log{\frac{\m}{\m_0}}\right)^{1/2}
\end{array}
\ee
The case of the scalar $A_5^n$, whose zero mode would play the role of the Higgs, is much
 more complicated. In any case one thing is
 clear: the correction is not finite even for the zero mode in the chiral theory $m_f=0$. 
In fact, since $A_5^0$ is massless at tree level we cannot absorb the divergence at one loop.
 For consistency of the theory one must include a mass term in the original lagrangian
\be
\mathcal{L}_m=\frac{1}{2}m_B^2A_MA^M\supset\frac{1}{2}m_B^2A_5^0A_5^0
\ee
But this is clearly non gauge-invariant (except precisely for the zero mode). 
Another possibility
 is to include a mass term only for the zero mode in the compactified Lagrangian 
but it would make
 the theory lose all the advantages of Gauge-Higgs unification coming from extra-dimensional
 gauge 
invariance and the problems associated with the mass of a scalar would reappear.
Nevertheless this interpretation is in sharp contrast with the (also four-dimensional) 
one in \cite{Cheng} where a 
totally finite result was obtained\footnote{Some authors \cite{Nilles, Kim} have found 
cuadratically divergent corrections with similar calculations, which suggests that this 
kind of computation is not very well established. There is also a quadratically divergent
Fayet--Iliopoulos term in supersymmetric theories on orbifolds \cite{Ghilencea2}.}.
 In particular the correction to the 
mass of the nonzero KK modes is found to be
\be
\delta m^2=-\frac{e^2\zeta(3)}{4\p^4}M^2
\ee
In the approximation $p^2=p_5^2$. The reason of the difference is of course the point 
where the sum over the extra-dimensional momentum is performed\footnote{In \cite{Cheng} a 
Poisson resummation is done before the proper time integral.}. 
Suppose we are trying to do a purely five-dimensional calculation of the diagram. 
Before the compactification of the theory, let us say to $\mathbb{R}^4\times S^1$, we have 
a full $O(1,4)$ invariance. In that case the momentum integral has trivially the property
\be
\int\frac{d^5 k}{(2\p)^5}f(k^2)=\int\frac{d^4k}{(2\p)^4}\int \frac{dk_5}{2\p}f(k^2)=
\int\frac{dk_5}{2\p}\int\frac{d^4k}{(2\p)^4}f(k^2)
\ee
Which means that it is strictly equivalent to perform the integral first over the 
extra dimension and then the four dimensional one or viceversa. If we now compactify the 
theory the full five-dimensional Lorentz invariance is spontaneously broken to $O(1,3)
\times O(2)$. An essential ambiguity \footnote{The ambiguity is related to considering 
$k_5$ as a component of the five-momentum, but usually it is treated as a mass for the 
higher KK modes. Then, it is natural to do the summation after the evaluation of a single 
diagram because in that case $k_5$ simply labels fermions with different masses.} 
appears then if we insist in interpreting the diagrams as five-dimensionals because clearly
\be\label{ineq}
\sum_{k_5}\int d^4k f(k_\m,k_5)\ne\int d^4k\sum_{k_5}f(k_\m,k_5)
\ee
When the integral (or the sum) is divergent. Those two alternatives are then the two
different four-dimensional calculations we were refering to above.
\par
This observation is not new and a lot 
of effort has been put into studying its possible consequences, also when the expresions 
are not formally divergent. In \cite{Delgado} a brane Gaussian distribution along the 
extra dimension was used to regularize the theory while KK modes were not truncated. 
The integral can be performed and after the infinite sum the result is claimed to be finite. 
Similar conclusions were reached in \cite{Contino} using Paulli-Villars and an adapted 
version of dimensional regularization. Both regulators are supposed to preserve the 
symmetries. The most explicit study of the validity of (\ref{ineq}) is that of 
\cite{Gambassi} were a method to dimensionally regularize KK sums using Mellin 
transform and analytic extension of special functions is proposed. With this procedure 
it is believed that the ambiguity is resolved. Works with a similar philosophy can be 
found in \cite{Groot}  where the tower is summed using a pole function and in 
\cite{DiClemente} were the sum is regularized using a $\zeta$-function.
In any case, we believe that none of these works is fully satisfactory.

\begin{thebibliography}{99}
\bibitem{Kim}
  H.~D.~Kim,
  ``To be (finite) or not to be, that is the question. Kaluza-Klein contribution to
 the Higgs mass,''
  arXiv:hep-ph/0106072.
\end{thebibliography}
\end{document}